\documentclass{article}
\usepackage{amsmath}
\usepackage{amssymb}
\usepackage{graphicx}
\hfuzz=\maxdimen
\begin{document}

\title{On the relation between boundary proposals and hidden symmetries of the extended pre-big bang quantum cosmology }
\author{S. Jalalzadeh\thanks{email: s-jalalzadeh@sbu.ac.ir},\,\, T. Rostami\thanks{email:t\_rostami@sbu.ac.ir}\\{\small Department of Physics, Shahid Beheshti University G. C., Evin, Tehran 19839, Iran}\,\, \and\,\, \\
P. V. Moniz\thanks{e-mail: pmoniz@ubi.pt}\\{\small Centro de Matem\'{a}tica e Aplica\c{c}\~{o}es- UBI, Covilh\~{a}, Portugal},\\
{\small Departmento de F\'{\i}sica, Universidade da Beira Interior, 6200 Covilh\~{a}, Portugal}}
\date{\today}
\maketitle
\begin{abstract}
A framework associating quantum cosmological  boundary conditions to
minisuperspace  hidden symmetries has been introduced in \cite{7}.
The scope of the application was, notwithstanding the novelty,
restrictive because it  lacked a discussion involving realistic
matter fields. Therefore, in the herein letter, we extend the
framework scope to encompass elements from a scalar-tensor theory
 in the presence of a cosmological constant. More precisely, it is
shown that hidden minisuperspace symmetries present in a  pre-big
bang model suggest a process from which boundary conditions can be
selected.

\end{abstract}

\section{Introduction}

\indent

Quantum geometrodynamics is the oldest and still active approach to
quantum gravity and quantum cosmology \cite{1}. Since it bears a
canonical set up in its foundation, it contains constraints as
central equations. In the case of a metric representation
perspective,  these are the Hamiltonian and the diffeomorphism
constraints and in the case of the connection (Ashtekar)  approach,
the Gauss constraints are also added.


Quantum geometrodynamics has, nevertheless, many technical and
conceptual challenges: the problem of time, the problem of
observables, factor ordering issues, the global structure of
spacetime manifold and the problem of boundary conditions (for more
details, see \cite{1}). The issue of boundary conditions for the
wave function of the Universe has been one of the most active areas
of quantum cosmology. Two leading lines have been the no-boundary
\cite{2} and the tunneling proposals \cite{3}. Two other proposals
(of a Dirichlet or a Neumann nature) have also been used, although
less often in the literature, to deal with the presence of classical
singularities. More precisely, the wave function should vanish at
the classical singularity (DeWitt boundary condition) \cite{4}, or
its derivative with respect to the scale factor vanishes at the
classical singularity \cite{5}.

All those boundary conditions are chosen {\it ad hoc}, with some
particular physical intuition in mind \cite{6}, possibly with some
characteristic symmetry element being imported to assist, but they
are not part of a dynamical law.
 Due to the fact that the algebra intrinsic to any given minisuperspace  is specified by the symmetries
 of the model, including the type of matter content, it may be of interest to investigate whether
 there is a relation between any set of allowed boundary conditions and the algebra associated to  the Dirac observables of the cosmological model.

Let us be more concrete. In \cite{7}, a simple closed
Friedmann-Lema\^itre-Robertson-Walker (FLRW) model in the presence of
either dust or radiation
 was studied. It was shown that by means of the presence of a hidden symmetry, namely $su(1,1)$, the model admits a particular
 Dirac observable, subsequently allowing to
 establish boundary proposals admissible for the model. More precisely,
 the Casimir operator $J^{2}=j(j-1)$ of the $su(1,1)$ algebra
 leads (for the value $-\frac{3}{16}$) to the discrete set $j=\{\frac{1}{4},\frac{3}{4}\}$ for the Bargmann index, which is  a gauge invariant observable.
 Then, it was shown that those two gauge invariant quantities split the underlying Hilbert space into two disjoint invariant
  subspaces, each corresponding to a different choice of boundary conditions (namely, of a Dirichlet or Neumann type).
  Notwithstanding the interest of this case study,
  it is a fact that it was a model of a very restrictive
 range.

 In what follows, we will extend the scope of that discussion and employ a scalar-tensor gravity
 theory.
 Scalar-tensor gravity theories seem to be relevant in explaining the very early universe as shown in \cite{8,9,10}. These
  theories are defined through a non-minimal coupling of a scalar field to the spacetime curvature, which is originated from the low energy limit
  of unified field theories such as superstring theory \cite{9,10,11}. With a suitable conformal transformation, theories with higher-order terms
  in the Ricci scalar may lead to scalar-tensor form \cite{12}. Moreover, higher-dimensional gravity leads to a scalar-tensor theory from a
  dimensional reduction \cite{13}. We make our analysis more concrete by employing a  spatially flat isotropic cosmology,
  in which the dilaton scalar field is non-minimally
  coupled to the spacetime curvature, in the
presence of a cosmological constant. There is a hidden symmetry,
which, leading to a particular Dirac observable,  provides a setting
where concrete boundary conditions can be  subsequently extracted
\cite{14}. This paper is then organized as follows. In section 2 we
present our model from a classical canonical perspective. In section
3 we formulate its quantization, briefly elaborating on the standard
canonical procedure as well as from a reduced phase space and
corresponding observables point of view. Furthermore, we describe
how from a hidden minisuperspace symmetry, taking into account the
observables algebra, concrete boundary conditions can be identified
within this framework. Section 4 contains a summary, a discussion as
well as an outlook on our results.

\section{The model}
\label{sec:1}

\indent

One of the simplest extension of Einstein gravity is the
scalar-tensor theory. The reduced string action in spatially flat
and homogenous cosmologies has a scale factor duality
\cite{15,16,17}. The symmetry group is ${Z_{2}}^{D-1}$, which
relates the expanding dimensions to contracting ones.   Scale factor
duality (SFD) is a special case of a more general $O(d,d)$ symmetry. It
has been shown \cite{18}, that the SFD at the
classical level is associated with a 
$N=2$ supersymmetry at quantum level \cite{14}.  In Ref. \cite{Red},
it is described that the concept of pre-big bang cosmology can be
extended beyond the truncated string effective action to include
more general dilaton-graviton systems.  It is interesting to study
this types of theories in their own right and they also place the
results and predictions of string cosmology in a wider scenery.

Let us start from a $D$-dimensional scalar-tensor theory, which is
non-minimally coupled to the spacetime curvature as
\begin{eqnarray}\label{2,1}
{\cal S}=\int d^{D}x\sqrt{-g}\ e^{-\phi}\left(R-\omega(\nabla\phi)^{2}-2\Lambda\right),
\end{eqnarray}
where $\cal R$ is Ricci scalar, $\phi$ is the dilaton field, which
plays the role of varying gravitational constant, g is the
determinant of spacetime metric, $\omega$ is a spacetime constant
and $\Lambda$ is the cosmological constant. When $\Lambda=0$, this
theory is equivalent to the standard Brans-Dicke theory. The
genus-zero effective action of the bosonic string reduces to the
above action when the antisymmetric tensor field $B_{\mu\nu}$
vanishes, $\omega=-1$ and $\Lambda=(D-26)/3{{\alpha}^\prime}$
(${\alpha}^\prime$, is the inverse string tension) \cite{11}. Also,
when $\Lambda$ is proportional to $(D-10)$, action (\ref{2,1})
represents the effective action for the bosonic sector of the closed
superstring.
Before proceeding, let us mention that a vast part of
the content of this section is a summarized extract of \cite{18}.

We employ a  spatially flat, isotropic and homogeneous FLRW model,
 parameterizing the metric by means of the line element
\begin{eqnarray}\label{2,2}
 {ds^{2}}=-N(t)^{2} dt^{2}+e^{2\alpha(t)} dx_{i}^{2},\,\,\,i=1,2,...,(D-1),
 \end{eqnarray}
where $N(t)$ is the lapse function and $e^{\alpha(t)}$ is the scale
factor of the universe. Then, the action reduces to
\begin{equation}\label{2,3}
{\cal{S}}=\int dt e^{(D-1)\alpha-\phi}[\frac{1}{N}(-(D-1)(D-2)\dot{\alpha}^{2}+
(D-1)\dot{\alpha}\dot{\phi}+\omega\dot{\phi}^{2})-2N\Lambda] \hspace{.15cm}.
\end{equation}
 SFD is a characteristic of our setting, allowing to discuss  string theory features within
 cosmology. In fact, the above action has SFD properties which are allowed by means of associating our analysis to a
spatially flat FLRW model \cite{18}. The action is symmetric under
the (SFD) simultaneous transformation
\begin{equation}\label{2,4}
\begin{array}{lll}
{\alpha}=\left[\frac{(D-2)+(D-1)\omega}{D+(D-1)\omega}\right]\tilde{\alpha}-\left[\frac{2(1+\omega)}{D+(D-1)\omega}\right]\tilde{\phi},\\
\phi=-\left[\frac{2(D-1)}{D+(D-1)\omega}\right]\tilde{\alpha}-\left[\frac{(D-2)+(D-1)\omega}{D+(D-1)\omega}\right]\tilde{\phi}.
\end{array}
\end{equation}
A conserved quantity can be identified,
$F=e^{\left((D-1)\dot{\alpha}-\phi\right)}\left[\dot{\alpha}+(1+\omega)\dot{\phi}\right]$,
as introduced in \cite{18}. The time reversal invariance of the
action under $t=-\tilde{t}$, in addition to the above transformation
(\ref{2,4}),  leaves $F$ unchanged. If we use the following
transformations
\begin{equation}\label{2,6}
\begin{cases}
{\ u}=2\epsilon^{\frac{1}{2}}\left[\frac{D-1+\left(D-2\right)\omega}{D+\left(D-1\right)\omega}\right]^{\frac{1}{2}}e^{\frac{1}{2}\left((D-1)\alpha-\phi\right)}
\sinh\left(\frac{\gamma}{2}\alpha+\frac{1}{2(D-2)}(\frac{D-1}{\gamma}-\gamma)\phi\right),\\
{\ v}=2\epsilon^{\frac{1}{2}}\left[\frac{D-1+\left(D-2\right)\omega}{D+\left(D-1\right)\omega}\right]^{\frac{1}{2}}e^{\frac{1}{2}\left((D-1)\alpha-\phi\right)}
\cosh\left(\frac{\gamma}{2}\alpha+\frac{1}{2(D-2)}(\frac{D-1}{\gamma}-\gamma)\phi\right),\\
\end{cases}
\end{equation}
where
\begin{equation}\label{2,15}
\begin{cases}
{\gamma}=\left[\frac{D-1}{D-1+\left(D-2\right)\omega}\right]^{\frac{1}{2}},\\
{\lambda}=-2\Lambda\left[\frac{D+\left(D-1\right)\omega}{D-1+\left(D-2\right)\omega}\right],\\
\epsilon= \pm 1.
\end{cases}
\end{equation}
then the action changes to the oscillator-ghost-oscillator form
\begin{equation}\label{2,7}
{ \cal S}=\frac{1}{\epsilon}\int dt\left[\frac{1}{N}\left(\dot{u}^{2}-\dot{v}^{2}\right)-\frac{\lambda}{4}\left(u^{2}-v^{2}\right)N\right].
\end{equation}
It is obvious that if $\epsilon=+1$, $\omega>\frac{-D}{D-1}$ and if
$\epsilon=-1$, $\omega<\frac{-D}{D-1}$. The minisuperspace signature
is $(+,-)$, therefore in the reduced action $u$ acts as the
``spacelike'' component and $v$ as the ``timelike'' component. In
the $\{u,v\}$ coordinate system, the duality symmetry is more apparent.
In fact,
 the invariance of the action under time reversal and  parity symmetry, which is introduced as
\begin{equation}\label{2,20}
 t\rightarrow -t,\hspace{.3cm} u\rightarrow -u,\hspace{.3cm} v\rightarrow
 v,
\end{equation}
leaves the conserved quantity
$F=\frac{\gamma}{2}\left(\dot{u}v-\dot{v}u\right)$, invariant. It
should be noted that the transformation (\ref{2,4}) can be seen
emerging from the above parity symmetry. From the action (\ref{2,7}) let us write its  Lagrangian:
\begin{equation}\label{2,8}
{\cal L}=\frac{1}{\epsilon}\left[\frac{1}{N}(\dot{u}^{2}-\dot{v}^{2})-\frac{\lambda N}{4}(u^{2}-v^{2})\right].
\end{equation}

In order to construct the Hamiltonian of the model, the momenta
conjugate to $u$, $v$ and $N$ are
\begin{equation}\label{2,9}
{\Pi_{u}}=\frac{2}{N\epsilon}\dot u,~~~~~{\Pi_{v}}=\frac{-2}{N\epsilon}\dot v,~~~~~~{\Pi_{N}}=0,
\end{equation}
which subsequently lead to
\begin{equation}\label{2,11}
{\cal{H}}=\frac{N\epsilon}{4}\left[(\Pi_{u}^{2}-\Pi_{v}^{2})+{\lambda}(u^{2}-v^{2})\right].
\end{equation}
The existence of the constraint $\Pi_N=0$ indicates that the Lagrangian of the system is singular and the Hamiltonian can be generalized by adding to it the primary constraints multiplied by arbitrary functions of time, $\zeta$. The total Hamiltonian will then be
\begin{equation}\label{2,12}
{\mathcal{H}_T}=\frac{N\epsilon}{4}(\Pi_{u}^{2}-\Pi_{v}^{2})+\frac{N\lambda\epsilon}{4}(u^{2}-v^{2})+\zeta{\Pi_{N}}.
\end{equation}
The constraint must be satisfied at all times and therefore,
\begin{equation}\label{2,13}
{\dot{\Pi}_{N}}=\{\Pi_{N},{\cal{H_{T}}}\}\approx0,
\end{equation}
which leads to the secondary (Hamiltonian) constraint
\begin{equation}\label{2,14}
H=\frac{N\epsilon}{4}(\Pi_{u}^{2}-\Pi_{v}^{2})+\frac{N\lambda\epsilon}{4}(u^{2}-v^{2})\approx0.
\end{equation}
The existence of the constraint (\ref{2,14}) means that there are
some degrees of freedom which are not physically relevant. Hence we
can fix the gauge as $N=$constant. Note that, by means of the
coordinate transformation $v=R\cosh\theta$ and $u=R\sinh\theta$, the
Hamiltonian (\ref{2,14}) becomes
\begin{equation}\label{x1}
H=\frac{N\epsilon}{4}\left(-\Pi_R^2+\frac{1}{R^2}\Pi_\theta^2-\lambda R^2\right),
\end{equation}
where $\Pi_R=-\frac{2}{N\epsilon}\dot R$ is the ``radial'' momentum
and $\Pi_\theta=\frac{2R^2}{N\epsilon}\dot \theta$  denotes a
conserved ``angular momentum''. It is easy to show that
$\Pi_\theta=\frac{2}{N\epsilon}(\dot uv-\dot
vu)=\frac{4}{N\epsilon\gamma}F$. Hence, the duality symmetry in
these new coordinates is equivalent to the anticlockwise
pseudo-rotation in time reversal; $\theta\rightarrow-\theta$,
$t\rightarrow-t$.

\section{ Reduced phase space quantization and Dirac observables}
\subsection{Standard quantization}\label{Standard quantization}

\indent

In this (most usual and straightforward procedure)  the quantization
of the system is made by replacing the canonical conjugate variables
$(u,\Pi_{u})$, $(v,\Pi_{v})$ by operators satisfying the commutation
relations $[x_{i},\Pi_{j}]=-i\delta_{ij} $. Thus, if we neglect any
ambiguities that may arise due to factor ordering, the Wheeler-De
Witt (WDW) equation can be written, from the Hamiltonian constraint
(\ref{2,14}), as
\begin{equation}\label{3,1}
\left[-\partial^{2}_{u}+\partial^{2}_{v}+\Omega^{2}(u^{2}-v^{2})\right]\Psi(u,v)=0,
\end{equation}
where $\Omega={\sqrt{\lambda}}$. The wave function of
the universe is easily obtained as
\begin{equation}\label{3,5}
\Psi_{n_{u},n_{v}}(u,v)={\mathcal N}H_{n_{u}}(\sqrt{\Omega}u)H_{n_{v}}(\sqrt{\Omega}v)e^{-\Omega(u^{2}+v^{2})/2},
\end{equation}
where $H_{n}$ is the Hermite polynomial of order $n$  and $\mathcal N$ is a
normalization constant. These solutions form a discrete basis for any bounded
wave function $\Psi=\sum c_{n}\Psi_{n}$, where $c_{n}$ are complex
coefficients. For the ground state, $n=0$, and $n>0$
correspond to the excited states. These states represents Euclidean
geometries, as they do not oscillate. Lorentzian geometries may be
obtained if the appropriate value for $c_{n}$ are taken
\cite{14,18}.

It should be noted that the classical solution has a singularity at
$u=0$ and $v=0$. In order to avoid this singularity, we can adopt
that the wave function vanishes at the classical singularity i.e., (De Witt
Boundary proposal)
\begin{equation}\label{3,2}
\Psi(u,v)|_{u=0,v=0}=0,
\end{equation}
or, as proposed by Tipler in \cite{19},
\begin{equation}\label{3,4}
\frac{d\Psi}{dx_{i}}|_{u=0,v=0}=0.
\end{equation}
 As is mentioned in \cite{20}, upon choosing any
of  the above boundary conditions, we obtain, for the
oscillator-ghost-oscillator system,
\begin{equation}\label{3,7}
n_{u}-n_{v}=0,
\end{equation}
which, for the  De Witt (\ref{3,2}) boundary condition, states that
both $n_{u}$ and $n_{v}$ must be odd, whereas  if the boundary
condition (\ref{3,4}) is taken, then both $n_{u}$ and $n_{v}$ must
be even.


\subsection{Reduced phase space and observables}

In the reduced phase space quantization, we first identify the
physical degrees of freedom of a given model at the classical level
by means of the factorization of the constraint surface with respect
to the action of the gauge group, generated by the constraints. Then
the resulting Hamiltonian system is quantized as a usual
unconstrained system \cite{RPS}.The constraint surface is obtained
by means of gauge transformations, generated by all the first class
constraints. A gauge invariant function on this surface is an
observable. A well known setting is  general relativity, which is
invariant under the group of spacetime diffeomorphisms and
consequently, the corresponding Hamiltonian can be expressed as a
sum of constraints \cite{1}. The point to take into consideration is
that from the associated Poisson bracket algebra we can describe the
classical dynamics of a system and any observable must commute with
these constraints.

With more detail, in order to find the gauge invariant observables
associated to the Lagrangian (\ref{2,8}), we consider the
unconstrained phase space $\Gamma$ in $\mathbb{R}^{4}$ with global
canonical coordinates $(x_{i},\Pi_{i})$. Subsequently, let us define
complex valued  holomorphic functions $S=\{C_{i}, C^{\ast}_{i},1\}$ on $\Gamma$
\begin{equation}\label{4,1}
\left\{\begin{array}{lll}
{\ C_{i}}=(\frac{1}{2\Omega})^{\frac{1}{2}}[\Omega x_{i}+i{\Pi_{i}}],\\
{\ C^{\ast}_{i}}=(\frac{1}{2\Omega})^{\frac{1}{2}}[\Omega x_{i}-i{\Pi_{i}}].
\end{array}
\right.
\end{equation}
These functions satisfy the Poisson brackets
$\{C_{i},C^{\ast}_{j}\}=-i\delta_{ij}$. The Hamiltonian constraint
can then be readily written as
\begin{equation}\label{4,2}
{\ H}=\Omega \left[C^{\ast}_{u}C_{u}-C^{\ast}_{v}C_{v}\right] .
\end{equation}

  Moreover, let us consider on $\Gamma$ the
following two sets of functions
\begin{equation}\label{4,4}
\tilde{J_{0}}=C^{\ast}_{u}C_{u}-C^{\ast}_{v}C_{v},
\end{equation}
and
\begin{equation}\label{4,5}
\left\{\begin{array}{lll}
J_{0}=\frac{1}{2}\left(C^{\ast}_{u}C_{u}+C^{\ast}_{v}C_{v}\right),\\
J_{+}=C^{\ast}_{u}C^{\ast}_{v},\\
J_{-}=C_{v}C_{u}.
\end{array}
\right.
\end{equation}
The second set of functions satisfy the following closed Poisson algebra
\begin{equation}\label{4,6}
\left\{J_{0},J_{\pm}\right\}=\mp iJ_{\pm},\hspace{.3cm}
\left\{J_{+},J_{-}\right\}=2iJ_{0}.
\end{equation}
Since the Poisson brackets of the above variables and the
Hamiltonian vanish,
$\{\tilde{J_{0}},H\}=\{H,J_{0}\}=\{H,J_{\pm}\}=0$,   their values on
the constraint surface are constants of motion. In addition,  the
phase space of the model is four dimensional, which implies that
there will be at most four independent constraints. The Hamiltonian
constraint (\ref{4,2}) implies $\tilde J_{0}=0$.  Furthermore, if we
define
\begin{eqnarray}\label{4,7}
J^{2}:=J_{0}^{2}- \frac{1}{2}\left(J_{+}J_{-}+J_{-}J_{+}\right),
\end{eqnarray}
then, by inserting definitions (\ref{4,5}) into the equation
(\ref{4,7}), we can easily show that on the constraint surface
$H=0$, the $J$'s are not algebraically independent but satisfy the
identity
 \begin{eqnarray}\label{aa1}
J^2=\tilde J^2_0-\frac{1}{4}=\left(\frac{H}{\Omega}\right)^2-\frac{1}{4}=-\frac{1}{4}.
\end{eqnarray}
Rewriting then the conserved
quantity $F$ in terms of these new set of variables, as
\begin{equation}\label{4,8}
F=\frac{i\epsilon\gamma}{4}\left(J_{+}-J_{-}\right),
\end{equation}
the vanishing of its Poisson bracket with the Hamiltonian is obvious.

\subsection{Reduced phase space quantization, hidden symmetries and boundary conditions}
\label{sec:3}

\indent

Regarding a quantum mechanical description, in order to determine
the wave function we must specify certain conditions at the boundary
of the system under consideration. However, in quantum cosmology
there is nothing external to the universe. It is assumed that an
independent physical law would define appropriate
 boundary conditions \cite{21}. Or, as we discuss herein, the symmetries of the cosmological model under investigation may suggest arguments for
 such selection. Indeed, using  hidden symmetries associated to the dynamics of the model will lead us to extract specific boundary conditions. More concretely,
 by means of considering the Dirac observables of the cosmological model.

Let us start by introducing the quantum counterparts of the set $S$
as $\hat{S}=\{C_{i},C^{\dag}_{i},1\}$, with the following commutator
algebra
\begin{equation}\label{5,1}
[C_{i},C^{\dag}_{j}]=\delta_{ij},~~~~~ [C_{i},1]=[C^{\dag}_{i},1]=0.
\end{equation}
The action of operators ${C_{i},{C^{\dag}}_{i}}$ on the physical Hilbert space is
\begin{equation}\label{5,2}
C_{i}|n_{i}\rangle=\sqrt{n_{i}}|n_{i}-1\rangle,\hspace{.2cm}
C_{i}^{\dag}|n_{i}\rangle=\sqrt{n_{i}+1}|n_{i}+1\rangle.
\end{equation}
To represent the Dirac observables of our model quantum
mechanically, we define  operators on the phase space. More
concretely, we notice that the (classical) Poisson bracket algebra
of the $u(1,1)$  for $J_{0}$ and $J_{\pm}$  and $\tilde{J_{0}}$
(operator of the $u(1)$ algebra) can be promoted into a commutator
algebra by setting the $u(1)$ generator as \cite{20}
\begin{equation}\label{5,3}
\tilde{J_{0}}=-C_{u}^{\dag}C_{u}+C_{v}^{\dag}C_{v},
\end{equation}
and the  generators of $SU(1,1)$ in two-mode realization as
\cite{two}
\begin{equation}\label{5,4}
\begin{cases}
J_{+}=C_{u}^{\dag}C^{\dag}_{v},\\
J_{-}=C_{v}C_{u},\\
J_{0}=\frac{1}{2}[C_{u}^{\dag}C_{u}+C_{v}^{\dag}C_{v}+1],
\end{cases}
\end{equation}
which  satisfy the following commutation relations
\begin{equation}\label{5,5}
\left[J_{+},J_{-}\right]=-2J_{0},\hspace{.3cm}
\left[J_{0},J_{\pm}\right]=\pm J_{\pm}.
\end{equation}
The above commutation relations represent the Lie algebra of
$su(1,1)$. The action of the above generators on a set of basis
eigenvectors $|j,m\rangle$, which are simultaneous eigenvectors of
$J_{0}$ and $J^{2}$, is given by
\begin{equation}\label{5,6}
\left\{\begin{array}{lll}
J_{0}|j,m\rangle=(j+m)|j,m\rangle,\\
J_{+}|j,m\rangle=\sqrt{(2j+m)(m+1)}|j,m+1\rangle,\\
J_{-}|j,m\rangle=\sqrt{m(2j+m-1)}|j,m-1\rangle.
\end{array}
\right.
\end{equation}
The positive discrete series of this Lie algebra are labeled by the
Bargmann index $j$, which is a positive real number $j>0$ and $m$ is
any nonnegative integer \cite{22}. According to (\ref{5,2}) and the
constraint $\tilde{J_{0}}=0$, we have for the $u(1)$ generator
\begin{equation}\label{5,7}
\tilde J_{0}|j,m\rangle=0,
\end{equation}
 Moreover, the
Casimir operator is defined as in \cite{23}
\begin{equation}\label{5,8}
\left\{\begin{array}{c}
J^{2}:=J_{0}^{2}-\frac{1}{2}(J_{+}J_{-}+J_{-}J_{+}),\\
J^{2}|j,m\rangle=j(j-1)|j,m\rangle,
\end{array}
\right.
\end{equation}
with the following commutation relations
\begin{equation}\label{5,9}
\left[J^{2},J_{0}\right]=0,\hspace{.3cm}
\left[J^{2},J_{\pm}\right]=0.
\end{equation}
Thus, the irreducible representation of $u(1,1)$ is determined by
the number $j$ and the eigenstates of $J^{2}$, $J_{0}$ and $\tilde
J_{0}$. Furthermore, the Hamiltonian can be written as
\begin{equation}\label{5,10}
H=\Omega\left(C^{\dag}_{u}C_{u}-C^{\dag}_{v}C_{v}\right)=-\Omega\tilde{J_{0}},
\end{equation}
which allow that the Casimir operator commutes with the Hamiltonian
\begin{equation}\label{5,11}
[J^{2}, H]=0,
\end{equation}
 which shows that the Bargmann index is a Dirac observable. In the
number operator representation, the basis states are $|n_v\rangle$
and $|n_u\rangle$, so that the Hilbert space of the ``two-mode''
field is the direct product
\begin{eqnarray}\label{aa2}
|n_v,n_u\rangle=|n_v\rangle\otimes|n_u\rangle.
\end{eqnarray}
However, it is desirable to map these direct product states on the
$\mathcal D^+(j)$ unitary irreducible representation of $SU(1,1)$.
To do this we first need to find the Bargmann index $j$. From the
realization of definition (\ref{5,4}), the Casimir operator can be
shown to be
\begin{eqnarray}\label{aa3}
J^2=\frac{1}{4}(\tilde J_0^2-1).
\end{eqnarray}
In general, according to the definition of $\tilde J_0$ in
(\ref{5,3}), the eigenvalues of this operator are just the
difference between $n_v$ and $n_u$. For  fixed eigenvalues of
$\tilde J_0$, with the condition $|n_v-n_u|\neq0$, using (\ref{5,3})
and (\ref{5,8}), the Bargmann index will be
\begin{eqnarray}\label{aa4}
j=\frac{1}{2}+\frac{1}{2}|n_v-n_u|=1,\frac{3}{2},2,...,\,\,\,\,\,|n_v-n_u|\neq0.
\end{eqnarray}
However,  in our model, the Hamiltonian constraint and equation
(\ref{5,10}) indicate that $n_v-n_u=0$. This means that the
Hamiltonian constraint forces  a degenerate case $n_v=n_u$ and
consequently, we get the unitary representation with a degenerate
Bargmann index
\begin{eqnarray}\label{aa6}
j=\frac{1}{2}.
\end{eqnarray}

Hence, the Bargmann index
$\{\frac{1}{2}\}$ is a gauge invariant observable of the quantum
cosmological model.
As $J^{2}$, $J_{0}$ and $J_{\pm}$ commute with the Hamiltonian, they
leave the physical Hilbert space $V_{H=0}$ invariant and consequently
we choose $\{J_{0},J^{2},\tilde{J_{0}},1\}$ as physical operators of
the model.  In addition, from (\ref{5,4}) we can easily examine how $J_{0}$
and $J_{\pm}$ act on $|n_{u},n_{v}\rangle$
\begin{equation}\label{5,12}
\begin{cases}
J_{0}|n_{u},n_{v}\rangle=\frac{1}{2}\left(n_{u}+n_{v}+1\right)|n_{u},n_{v}\rangle,\\
J_{+}|n_{u},n_{v}\rangle=\sqrt{(n_{u}+1)(n_{v}+1)}|n_{u}+1,n_{v}+1\rangle,\\
J_{-}|n_{u},n_{v}\rangle=\sqrt{n_{u}n_{v}}|n_{u}-1,n_{v}-1\rangle.
\end{cases}
\end{equation}
Comparing equations (\ref{5,12}) with equations (\ref{5,6}) and using $j=1/2$,
we obtain
\begin{equation}\label{5,13}
n_{u}=n_{v}=m.
\end{equation}
 According to the equation (\ref{5,5}), $J_{+}$ and $J_{-}$ do not commute
with each other, so they are not mutually compatible observables. It
is therefore impossible to simultaneously measure $J_{+}$ and
$J_{-}$. This means that odd $m$'s are
separated from even $m$'s. For example, consider that $m$ is an odd
number, then $m+1$ is even, so, if we measure an odd $m$, then the
information about $m+1$ (which is even) is not accessible.

 Let us now investigate in more
depth how the presence of the duality symmetry in our quantum
cosmological model can point to concrete boundary conditions to
consider. We remind from subsection \ref{Standard quantization} that
a wave function of the universe can be constructed admitting the De
Witt's boundary proposal (\ref{3,2}) or  the boundary proposal
(\ref{3,4}). We now aim to establish, for the herein found Dirac
observables and algebraic features, how symmetry considerations will
assist us in identifying a set of such conditions to chose one from.
We therefore represent the time reversal mentioned in section
\ref{sec:1} concerning  action (\ref{2,3}), by means of introducing
 the time reversal operator $\Theta$,  $\Theta\Pi_i\Theta^{-1}=-\Pi_i$. The behavior of
the $u(1,1)$ operators under this time reversal operator is
\begin{equation}\label{5,15}
\begin{array}{lll}
\Theta \tilde{J_{0}}\Theta^{-1}=\tilde{J_{0}}, ~~~~~\Theta J_{0}\Theta^{-1}=J_{0},\\
\Theta J^{2}\Theta^{-1}=J^{2},~~~~~ \Theta J_{\pm}\Theta^{-1}=J_{\pm}.
\end{array}
\end{equation}
In order to further illustrate the transformation (\ref{2,4}), or
equivalently the simultaneous change of $u\rightarrow -u$ and
$v\rightarrow v$, we introduce the parity operator in minisuperspace as $\pi^\dagger u\pi=-u$. The
parity operator acts then on the $u(1,1)$ operators as
\begin{equation}\label{5,16}
\begin{array}{lll}
\pi^{\dagger}\tilde{J_{0}}\pi=\tilde{J_{0}}, ~~~~~\pi^{\dagger} J_{0}\pi=J_{0},\\
\pi^{\dagger}J^{2}\pi=J^{2},~~~~~ \pi^{\dagger} J_{\pm}\pi=-J_{\pm}.
\end{array}
\end{equation}
Hence, $F$ respects the symmetry (cf. eq. (\ref{2,20}))
\begin{equation}\label{5,14}
\pi^{\dagger}\Theta F \Theta^{-1}\pi=F.
\end{equation}
The symmetry underlying the
transformation (\ref{2,20}) is equivalently represented into the
action of the operator $\pi^{\dagger}\Theta$. Using (\ref{5,10}),
(\ref{5,15}) and (\ref{5,16}), this allows to re-affirm that $H$ is
indeed invariant under the SFD and time reversal,
herein expressed in terms of $\pi^\dagger \Theta$, in correspondence
to the classical description
 in section \ref{sec:1}. The relevant feature to stress is that as $F$ commutes with the
Hamiltonian, $F$ is therefore also a gauge invariant observable .

At this point, having presented the necessary framework as well as
essential elements for the argument, let us explain how from the
hidden minisuperspace symmetries we can suggest a process from which
boundary conditions can be chosen. For that purpose, to make
concrete a physical realization of the  SFD together with the time
reversal, we  consider the pre-big bang cosmological scenario within
the context of the scalar-tensor theory of gravity \cite{Red}. The
aim in any version of that scenario, which for convenience we are
adopting to apply our framework, the universe starts out in a
contracting pre-big bang phase, would then goes through a bounce and
finally it emerges as an expanding post-big bang universe. Hence,
the bounce is represented by the self-dual point in the SFD.
However, as the literature and research work has provided, this has
not been fully achieved in terms of an  effective description or
workable SFD cosmology: these two phases are separated by a
curvature singularity. Assuming, nevertheless,  that the singularity
may disappear by including quantum gravitational or higher-order
corrections  from string theory, we can identify, from the duality
invariance of the equations of cosmological model, a  clear
suggestion about a possible temporal completion, based on a
``self-duality principle'' \cite{dual}.

An approach and methodology  to achieve an exit from the pre-big
bang phase, $(t<0)$, as described in the above paragraph, could be
provided by  quantum cosmology \cite{qcosmology}. If we indeed
assume that quantum effects would eventually remove the curvature
singularity, we can expect that there exist a smooth wave function
for the whole of the universe, including pre- and post- big bang
phases properly matched together.

Being therefore more concrete, by applying this context to the
simple setting studied here, introduced and described in detail in
the previous sections, let us take that the wave function of the
universe  $\Psi_n(u,v)$ in (\ref{3,5}) encompasses the phases of the
pre-big bang with  $(t'=-t,u'=-u,v'=v)$ and post big bang with
$(t,u,v)$. We now concentrate on the discrete Dirac observables
 of model.  According to  equation (\ref{5,16}), $[H,\pi]=0$. Consequently, $\Psi(u,v)$
is also an eigenfunction of parity operator, with eigenvalues $\pm
1$ which are Dirac observables. Similarly, the Hamiltonian commutes
with the time reversal,  $[H,\Theta]=0$, which, together with the
Hamiltonian constraint $H=0$ [or equivalently $n_v=n_u$ (the
eigenfunctions are nondegenerate)] implies that the wave function is
real, $\Psi_n(u,v)=\Psi_n^*(u,v)$. Hence, a general wave function is
given by $\sum_{n \,\text{ even}}c_n\Psi_n$ (even parity) or
$\sum_{n \,\text{odd}}c_n\Psi_n$ (odd parity), with real
coefficients. Therefore, the states of the Hilbert space can be
classified in terms of these two values of the parity operator, as
\begin{eqnarray}\label{aa8} V_{H=0}=V_{H=0,\pi=+1}\oplus
V_{H=0,\pi=-1}.
\end{eqnarray}
Thus, the gauge invariance of the parity implies a partition of the
Hilbert space into two disjointed invariant subspaces, which are
equivalent to the result of imposing boundary conditions (\ref{3,2})
or (\ref{3,4}), respectively.

\section{Conclusions,  Discussion and Outlook}
\label{sec:4}

\indent

 In this letter, we extended the scope of \cite{7} to obtain
relevant boundary conditions for a quantum cosmological model, by
means of identifying the corresponding necessary Dirac observables.
In more detail, we considered a non-minimally coupled scalar field
in a FLRW universe with a  cosmological constant in the context of a
scalar-tensor cosmology. More specifically, in order to
 include the SFD of the pre-big bang setting  extracted from string
 theory features, we considered a
spatially flat universe.

The reduced phase space quantization was then investigated. The
irreducible operators of the Lie algebra $u(1,1)$ were shown to have
a vanishing commutation relation with the Hamiltonian. From the
vanishing of the commutator of the $su(1,1)$ generator with the
Hamiltonian, together with  the gauge invariance of the Bargmann
index, this fixes the allowed states for the wave function of the
universe. Let us be more clear and elaborate more broadly. In Ref.
\cite{7} a closed FLRW universe filled with either dust or radiation
was considered, which was discussed by means of a Hamiltonian and
pointed to be equivalent to a one dimensional simple harmonic
oscillator. The hidden symmetry of that model was $su(1,1)$, with
the set of gauge invariant Bargmann values
$\{\frac{1}{4},\frac{3}{4}\}$, which is related to the so called
Barut-Girardello (even-odd coherent) states \cite{30}. This led to a
split in the underlying Hilbert space into two disjoint invariant
subspaces, each then subsequently shown to be corresponding to
different choice of boundary conditions, as (\ref{3,2}) and
(\ref{3,4}) to be more precise.

In a herein similar procedure (but bearing intrinsic differences
with respect to some elements in \cite{7}),
 we extracted  the hidden symmetries of the cosmological
scenario in study, which lead us to the set of Dirac observables of
model. The presence of a non-minimally scalar field as the matter
component in a spatially flat universe conveyed us to extended
symmetries, namely $u(1,1)$ together with time reversal (with
respect to the comoving time) and parity in minisuperspace.  The
Hamiltonian of the model studied in this paper is (regarding that in
\cite{7}) instead equivalent to the oscillator-ghost-oscillator
system, which leads here to a two-mode realization of the $su(1,1)$
algebra. This specificity of our setup induced that the Hamiltonian
constraint implies now a degenerate Bargmann index for model. Hence,
unlike to  ref. \cite{7}, the continuous symmetries of our herein
model are not the responsible for pointing to the boundary
conditions. More concretely, it is instead  the scale factor duality
of the cosmological model plus time reversal, which are equivalent
to the operator $\pi^{\dag}\Theta$ action, that allow here to select
specify the boundary conditions, associated to  the partition
(\ref{aa8}).

Notwithstanding the contribution we think this approach and
framework brings to quantum cosmology, there are issues where
additional more work is needed and indeed constitute new lines to
explore.

On the one hand, by employing a homogenous (and isotropic) model we
are neglecting an infinity of degrees of freedom, namely the
inhomogeneous modes that a wider metric or matter fields would
provide; Only the presence of these latter ones would bring a more
substantial realistic sense to  this methodology. The scope of
application of the framework is therefore still very restrictive, in
spite of this paper being a development with respect to the scarce
content in \cite{7}. Extending the herein framework to either FLRW
models, where homogeneity and isotropy would be the background, with
fluctuations or even some inhomogeneous simple models (e.g., Gowdy)
is needed to test it and establish if the scope can reach a wider
domain of cosmological models, taking into consideration the
intrinsic symmetries of the corresponding minisuperpaces. In
addition, the extension to (i) anisotropic homogenous cosmologies,
as in \cite{Clancy:1998ka}, or (ii) wider string settings as in
\cite{Red}, \cite{Cavaglia:2000rx}, taking the symmetries therein as
necessary ingredients, constitute tentative routes to consider.

On the other hand, the relation between SFD and other symmetries has
been pointed in the past (see e.g., \cite{18}, \cite{Moniz:2000uh}
and widely elaborated in \cite{14}), namely supersymmetry, therefore
including anti-commuting variables in the corresponding
minisuperspace configuration or phase space, by means or taking at
the start an action based, e.g., on a (albeit simplified)
supergravity setting. We can say that our guiding target is to
establish a robust correspondence involving minisuperspace
symmetries (for bosonic and fermionic degrees of freedom) and
subsequently allowing to specify how and which boundary conditions
can be suggested to select. This has not yet been done. Enlarging
the herein scope so that boundary conditions could be conveyed from
fundamental symmetries such as a supersymmetric type, would be
interesting. In addition, we can consider to include settings where,
besides the usual (commuting) space-time variables, non-communting
variables and deformed Poisson algebra \cite{Jalalzadeh:2014jea} are
present, in order to investigate the limits of applicability of this
framework, so far discussed within simple minisuperspace models.
Finally, factor ordering issues that emerge in the traditional
direct canonical quantization should also be considered and conveyed
into the discussion, as means of both enlarging and testing the
range of use of this framework.


\textbf{Acknowledgments} This work was in part supported by the
grant PEst-OE/MAT/UI0212/2014.


\end{document}